\documentstyle[pra,aps,epsf,epsfig,epic,twocolumn]{revtex}
\tighten

\begin{document}

\title{On the stability of long-range sound propagation 
through a structured ocean}

\author{Michael A. Wolfson and Steven Tomsovic}

\address{Department of Physics, Washington State University, 
Pullman, WA 99164-2814.}

\date{\today}

\maketitle
\noindent
\begin{center}
\end{center}

\begin{abstract}
Several acoustic experiments show a surprising 
degree of stability in wave fronts propagating 
over multi-megameter ranges through the ocean's 
sound channel despite the presence of random-like, 
sound speed fluctuations.  Previous works have 
pointed out the existence of chaos in simplified 
ray models incorporating structure inspired by 
the true ocean environment.  A 
``predictability horizon'' has been introduced 
beyond which stable wave fronts cease to exist 
and point-wise, detailed comparisons 
between even the most sophisticated models and 
experiment may be limited for fundamental reasons.  
We find, by applying one of the simplified models, 
that for finite ranges, the 
fluctuations of the ray stabilities are very 
broad and consistent with lognormal densities.  
A fraction of the rays retain a much more stable 
character than the typical ray.  This may be one 
of several possible mechanisms leading to greater 
than anticipated sound field stability.  The 
lognormal ray stability density may underlie 
the recent, experimentally determined, lognormal 
density of wave field intensities [J. Acoust. Soc. Am. {\bf 105},
3202--3218 (1999)].  
\end{abstract}


\section{Introduction}
\label{intro}

There is a great deal of experimental and theoretical 
interest in long-range, low-frequency acoustic pulse propagation 
through the deep ocean's sound channel.  It has 
been investigated as a problem of wave propagation in random
media (WPRM)~\cite{catoc,uscinski}, and as a basis for 
tomography~\cite{munk_oat,atoc98}.  
Recent results from the Acoustic Engineering Test (AET) 
as part of the Acoustic Thermometry of Ocean Climate
(ATOC) project can be found in Colosi {\it et al.}~\cite{colosi99}
and Worcester {\it et al.}~\cite{worcester99}.  
One of the main challenges in analyzing and 
understanding long range acoustic propagation is in 
dealing with difficulties arising from the ocean 
environment's tendency to generate multiple, weak, 
small-angle (forward) scattering~\cite{flatte_book}. At sufficiently 
long ranges of propagation, the multiple scattering 
should effectively randomize an acoustic pulse so that
it is very difficult to deduce much information.  However, 
several long range experiments have found a great deal 
of stability in the earlier portions of the received 
wave fronts in spite of the fluctuations inherent in the ocean 
environment~\cite{colosi94,spiesberger94,atoc98}.  
In addition, it has been found that wave field intensity 
fluctuations at long range are consistent with a lognormal 
density~\cite{colosi99} which would be reminiscent of earlier 
work in optics on WPRM~\cite{wprm}, except that this earlier work
was for the short range (weak focusing) regime.

In the past 10-15 years, simplified models inspired by 
the ocean environment have been shown to possess chaotic 
ray limits~\cite{tappert2,smith1,smith2}.  
Essentially simultaneously, there has been enormous progress 
in the understanding of chaotic systems~\cite{chaos}.  Some 
of the most familiar emerging concepts are simpler 
for bounded systems and are not easily applicable 
to open, scattering systems as we have here.  However, there 
is an important tool which does straightforwardly 
generalize for our 
purposes, the stability analysis of the rays.  
Stability matrices can be constructed as a function 
of range for each ray.  Their properties, such as the stability 
exponents, reveal the basic character of the 
rays, and are at the foundation of the findings 
reported in this paper.

There are several intriguing questions that arise from 
comparing the theoretical results to date regarding 
chaotic acoustic ray dynamics in the ocean and the 
high amount of stability observed in the data.  The 
most general question concerns how an acoustic pulse 
-- which at multi-megameter ranges extends to nearly 10 
seconds in time and 2 km in depth -- loses it's 
coherence from multiple forward scattering through 
interaction with internal waves and mesoscale 
energetics.  Because refraction is adequate to explain 
the scattering physics~\cite{simmen}, the 
ray limit should suffice for 
understanding long-range propagation.  
Some manifestations of the
underlying chaotic dynamics should be observed.

It has been suggested that there exists a 
``predictability horizon'' at the range of 
propagation defined by the scale over which 
chaotic dynamics develops~\cite{wolfson2}.  Beyond 
this range, the wave fields should appear as 
random superpositions of many plane waves~\cite{berry,berry2} 
which would imply that acoustic field intensity fluctuations 
are Rayleigh distributed~\cite{rayfluc,rayfluc2}.  
Several problems crop up beyond the predictability horizon.  It 
becomes increasingly difficult to get 
numerically calculated rays to converge to true 
rays of the system.  Worse, semiclassical approximations
(i.e.~wave front reconstructions) from the rays might  
fail for fundamental reasons related to the 
breakdown of stationary phase approximations,
but one should recognize that more optimistic
viewpoints exist on this issue~\cite{tomsovic,sepulveda,brown}.
Whether or not this is true, it is currently not
known to what extent tomographic inversions fail for any 
system beyond its predictability horizon where 
eigenrays are proliferating exponentially fast 
with increasing range.  
In order to begin addressing these and related issues,
we focus on the `forward propagation 
problem' by performing a statistical analysis that should 
be much less sensitive to the difficulties engendered 
by the predictability horizon.  

In fact, justifications for statistical laws derived by 
invoking stochastic or ergodic postulates are often ultimately
founded on the presence of fully developed chaos; see 
for example Ref.~\cite{prm}.  Systems 
that once were approached by stochastic methods have 
more recently begun to be regarded from the perspective  
of dynamical systems.  The two approaches mostly give 
consistent results, but there are important distinctions.   
Stochastic ray modeling 
is the traditional approach to the geometric limit of
the problem of WPRM~\cite{chernov}.  
This nondeterministic treatment leads one 
to pessimistic conclusions regarding the overall 
stability expected in an ocean acoustic pulse
at sufficiently long range~\cite{wolfson2}.  
By carefully defining the Lyapunov exponent, it turns out 
to be roughly half the value reported in Ref.~\cite{wolfson2}.  
The scales relevant to the ocean are such that 
this factor two increase in an important length 
scale might prove to be significant.  
Also, for this problem, the validity of the stochastic 
or ergodic assumptions deserves to be critically examined.  
It is not obvious that a dynamical systems perspective 
would lead to similar pessimistic conclusions as 
does the stochastic ray theory.  For example, 
the predictability horizon concept that has grown 
out of the chaotic dynamics point of view does not 
necessarily lead to a sudden transition --- regular behavior at 
short ranges, completely stochastic just beyond --- and remnants 
of stability that violate assumptions of stochasticity could 
persist well into the horizon's initial onset.  
We anticipate
several features of deterministic dynamics playing an
interconnected role in this regard, but 
we focus on the importance of only one, 
{\it intermittent-like dynamics}.  Intermittency 
is a common feature for nonintegrable dynamical 
systems~\cite{zisook,berge}.
For the ray acoustics problem, intermittent-like behavior 
is evident through the appearance of rays which persist in remaining
relatively insensitive to their initial conditions (also environment)
for remarkably long ranges, as measured on the inverse scale of the
mean Lyapunov exponent. 
It might then be expected that the existence of 
intermittent-like dynamics might allow
linear based tomographic inversions based on acoustic ray 
models to be suitable to greater ranges than previously 
anticipated.  

One objective of this article is to illustrate the existence of 
intermittent-like dynamics in the generic long-range
ocean acoustics problem.  This is a direct consequence
of the wide variability in the eigenvalues of the
stability matrix which is defined in Sect.~\ref{anal}A.
We demonstrate that the magnitude of its
largest eigenvalue follows a lognormal distribution, and that
the stability exponent follows a Gaussian distribution. 
Importantly, there is preliminary evidence suggesting that 
these distributions are robust, i.e.~that they would be found 
in much more realistic, sophisticated ocean models~\cite{colosi2}.  
To be more explicit,
if one knows the probability density of the stability exponents, 
then one can determine the expected measure of 
intermittent-like rays that will persist out to the reception range.  
It follows that these rays will 
not require extremely precise numerical interpolation schemes for 
quantities such as the gradient of the potential.

The model upon which we rely in this paper is admittedly 
extremely simplified.  However, it is not the model that is 
of concern, it is whether or not general features of simplified 
WPRM models carry over to the ocean itself.  If we are 
careful enough, the simplifications that we accept remove 
non-essential complications for uncovering the general 
physical features of interest, and no more.  A follow-up 
study to this one is underway which
uses a more realistic ocean sound speed model.
It is important for confirming the applicability of
our results to long-range ocean acoustics experiments.

The organization is as follows:  in Sect.~\ref{model}, 
we introduce and motivate a simple model leading to a 
one-degree-of-freedom, non-autonomous 
Hamiltonian dynamical system for the rays.  This is 
followed by a discussion of the analysis methods which 
are most critical for our study.  They are based on the 
stability matrix and its well known properties.  
Sect.~\ref{fluc} examines the fluctuation behavior of 
the stability exponents giving their densities as a 
function of range.  The proportion of intermittent-like 
rays is deduced and compared with the numerical results 
of the model.  We finish with a discussion and 
conclusions.  

\section{From wave equation to ray model}
\label{model}

We briefly outline the assumptions and approximations 
leading to the highly idealized ray model used in this paper. 
The primary physics we are concerned about involve refraction
of acoustic energy due to volume inhomogeneities in the ocean
sound speed.  We assume that interactions of the acoustic
energy with both the surface and sub-bottom are negligible.  
For multi-megameter ranges of propagation in mid-latitude, deep ocean
environments, a significant amount of acoustic energy is received
that satisfies this assumption \cite{munk_oat}.
As alluded to above, the necessary
assumptions leading to the primary results are that: 
i) the linear, one-way Helmholtz wave equation is valid 
(the important point here is that backscattering is negligible), and 
ii) the spatial scales of the sound speed field are long 
compared to the acoustic wavelength so that ray theory is
justified.  A detailed derivation is readily available
\cite{fred}.  We point out {\em a priori} that the coordinate
system is three-dimensional Cartesian ${\bf x}=(r,y,z)$, with $r$ the
range from the source, $y$ the transverse or cross-range coordinate, and
$z$ the depth from the surface. Thus, Earth curvature 
effects are neglected.

The fundamental starting point is the linear acoustic wave equation
\cite{flatte_book}:
\begin{equation}
\label{wave}
{1\over c^2}{\partial^2 \psi({\bf x};t) \over 
\partial t^2} = \nabla^2 \psi({\bf x};t) \; ,
\end{equation}
\noindent where $\psi({\bf x};t)$ is the complex scalar wave function
whose real part denotes the acoustic pressure. 
The sound speed field $c$ can be taken as 
a function of space ${\bf x}$ only
whereby it has been assumed that the time scales for the propagation
of the acoustic wave function are small compared to 
the time scale associated with the evolution of the sound speed field.
For non-dispersive sources, the acoustic group and phase
speeds are equivalent, and one can linearly transform 
Eq.~(\ref{wave}) from time to frequency, arriving at a Helmholtz
equation with the magnitude of the wave vector defined 
as $k=2\pi f/c$, where $f$ is the continuous wave (CW) source frequency.  
Attenuation effects can of course be incorporated
by modifying $k$ to be a complex quantity, but since we are interested
in: 1) acoustic energy that interacts negligibly with the ocean bottom,
and 2) typical sources operating at frequencies with minimal volume attenuation
(with center frequencies of about 100 Hz\cite{worcester99}), 
ignoring attenuation effects seems reasonable.
Also, one can similarly derive  
a reduced wave equation which includes variations in density,
but we ignore this effect because
it is known to be important predominantly with acoustic energy 
that interacts with the ocean sub-bottom, which is not considered herein.

The next assumption (which is quite a reasonable one) is that the strength
of the sound speed fluctuations, whatever the physical process that 
produces them,
are small.  This allows one to neglect backscattered acoustic energy,
and admits the one-way Helmholtz wave equation,
whereby one assumes a primary direction
of propagation along the range. 
The so-called 
`` standard parabolic approximation'' is invoked next. 
This allows
one to derive a linear partial differential wave equation of parabolic type
for the complex envelope of $\psi$.  The principle assumption is 
that this envelope wave function evolves slowly on the scale of 
the acoustic wavelength.
There many {\em flavors} of parabolic approximations that 
have varying degrees of phase errors in the complex wave function $\psi$
as compared to the one-way Helmholtz
equation \cite{brown_phase}, but we choose to use the standard 
parabolic approximation, which takes the form
\begin{equation}
\label{schro}
-{i\over k_0}{\partial \phi(y,z;r) \over \partial r} = {1\over k_0^2}
\nabla_{\bot}^2 \phi(y,z;r) + V(y,z;r)\phi(y,z;r) \; , 
\end{equation}
where the transverse Laplacian is represented by 
$\nabla_{\bot}^2=\partial^2_{y}+\partial^2_{z}$,
and the variable $r$ is the range (propagation variable), but 
plays an exact analogous role to time in the Schr\"odinger 
equation of quantum mechanics.  The parameter $k_0=2\pi f/c_0$ 
represents the reference wave number, and depends on the choice of
a reference sound speed $c_0$, which we take to be 1.5 km/s.
The potential, $V(y,z;r)$, is 
related to the sound speed fluctuations as
\begin{equation}
\label{potential}
V(y,z;r)=\frac{1}{2}\left [ \left (\frac{c_0}{c(y,z;r)}\right )^2 -1\right ] 
\sim {\delta c(y,z;r)\over c_0}\; ,
\end{equation}
where the sound speed variations away from an average 
profile has been expressed as $c(y,z;r)= c_0+\delta c(y,z;r)$.  
The sound speed fluctuations refract the rays 
and lead to chaos in a deterministic, mathematically defined 
sense.  Under the parabolic approximation, the basic 
problem maps precisely onto problems of quantum 
chaos~\cite{tomhel}.  The fields of long range 
acoustic propagation in the ocean and quantum chaos 
thus have the opportunity of cross-fertilization.  

Because the instability does not critically depend 
on having multiple degrees of freedom, we make a 
significant, practical simplification in the model of 
ignoring the depth degree of freedom ($z$); see 
Ref.~\cite{wolfson2} for a more detailed discussion of 
the model presented here.  The system could be thought of as lying in 
the plane of the sound channel axis, but this is really just
the generic problem of WPRM (see, for example, \cite{uscinski}).  
The gain in simplicity more than compensates for 
the loss of realism at this point as long as 
the main physical phenomena carry over to more 
realistic models.  As was mentioned in the 
Introduction, preliminary evidence for our main 
results have been found in recent calculations 
incorporating a much more realistic model~\cite{colosi2}.

The magnitude of the wave vector $k$ is large enough that for the 
purposes of this study, we can focus on the ray limit.  
The rays can be generated by a system 
of Hamilton's equations
\begin{eqnarray}
\label{hamilton}
\frac{dy}{dr} & = & \frac{\partial H(y,p;r)}{\partial p}  \; ,
\nonumber \\
\frac{dp}{dr} & = & -\frac{\partial H(y,p;r)}{\partial y}  \; ,
\end{eqnarray}
where $y$ and $p$ are the phase space variables 
cross-range (position) and horizontal slowness 
(momentum) respectively.  The independent variable $r$
denotes range.  Correspondence with Eq.~(\ref{schro}) necessitates
that the Hamiltonian is explicitly
\begin{equation}
\label{hamiltonian}
H=\frac{p^2}{2} + V(y;r) \; .
\end{equation}
The physical meaning of momentum is $p=\tan{\theta}$, 
where $\theta$ represents the angle a ray
subtends in cross-range about the range axis.  

\begin{figure}
\begin{center}
\leavevmode
\epsfxsize = 8.5 cm
\epsfbox{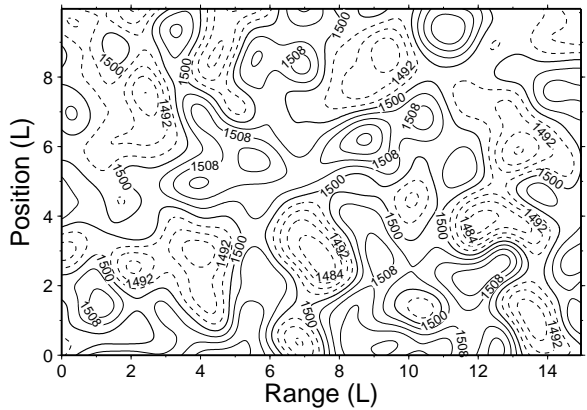}
\end{center}
\caption{
A portion of a realization of $V(y;r)$ whose 
full domain is $ 20 \mbox{L}
\times 320 \mbox{L}$ in $y$ and $r$ respectively.
The realization is constructed by the method described in 
Ref.~16.  Contours are labeled in units of sound speed (m/s).
The heavy solid contour line indicates the reference sound speed of 
1500 m/s.  The normalized root-mean-square fluctuations is 
$\epsilon =5 \times 10^{-3}$.}
\label{vyr}
\end{figure}

\begin{figure}
\begin{center}
\leavevmode
\epsfxsize = 7.5 cm
\epsfbox{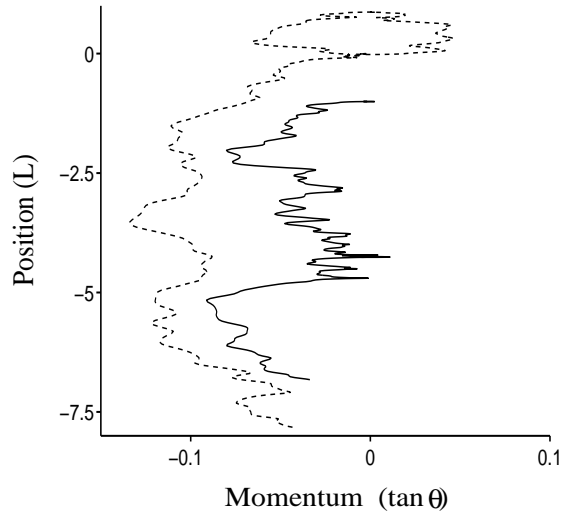}
\end{center}
\caption{Two distinct ray trajectories 
which have travelled through the sound speed field characterized in 
Fig.~\ref{vyr} are shown out to the range of 160 L.
The trajectory which starts at zero position and
zero momentum (dashed line) is highly unstable, 
and the trajectory which starts at unit position and zero momentum 
(solid line) is stable.}
\label{random}
\end{figure}

The state of the ocean is constantly changing, and its exact 
state is unknown.  A statistical ansatz is thus 
fruitful for making assertions concerning its ``typical'' 
state.  Assuming isotropy in the 
sound speed fluctuations in range and cross-range, the 
potential is taken to be a realization of a zero-mean, 
stationary, random function.  Thus a single 
correlation length scale $L$ exists. The standard deviation 
is denoted by $\epsilon = c_0^{-1}\langle \delta c^2 \rangle ^{1/2}$, where 
$ \langle \delta c^2 \rangle^{1/2}$ is the root-mean-square 
fluctuation of the sound speed about $c_0$.  
Typical values for underwater acoustics are
$\epsilon=O(10^{-3})$, 
and $L=O(100) \; \mbox{km}$, but both $\epsilon$ and $L$ vary
plus or minus an order of magnitude depending on what ocean structure
is considered and the geographic location.
For purposes of studying a fully defined, deterministic 
dynamical system, we complete
the description of $V$ by defining its correlation function to be
Gaussian,
\begin{equation}
\label{gaussian_pot}
\langle V(y;r) V(y+{\delta y};r+{\delta r}) \rangle = 
\epsilon^2 \exp{[-({\delta y}^2 + {\delta r}^2)/L^2]} \; .
\end{equation}
We exploit this single scale throughout the rest of this
article by transforming space variables
as $r \rightarrow r/L$ and $y \rightarrow y/L$, so the
physical dimensions will always be in units of $L$.  
One should envision the potential as 
being deterministic, even though it is selected from 
an ensemble of realizations.  This implies that the potential 
is to be considered a highly complicated (albeit smooth and fixed) 
function of both $y$ and $r$.  
To provide some idea of the character of this potential,
contours of sound speed fluctuations based on
a typical region of $V(y;r)$ is 
shown contoured in Fig.~\ref{vyr}.  The boundary conditions 
are taken as open in $y$, but numerically $y$ is treated as
periodic, with the ray coordinate unfolded {\em a posteriori}
to simulate the open boundary condition. A variety of initial
conditions are possible with the restriction that the initial momentum
is always kept small enough that the parabolic approximation is valid
all along the rays.  The rays deriving from two such 
initial conditions are plotted in Fig.~\ref{random} 
which shows their phase space portraits (position, momentum).  
In the absence of a varying potential, the solutions 
to the equations of motion are $p(t)=p_0$, $q(t)=p_0 t+q_0$.  
In this figure, rays would trace out vertical lines 
except in the case, $p_0=0$, in which rays would show up 
as points, $(q(t),p(t))=(q_0,0)$.  With the potential 
included, the rays trace out a random-walk-like motion 
with some drift as they move further away from the 
$p=0$ line. 


\section{Analysis Methods}
\label{anal}

The standard analysis of ray stability in the theory 
of dynamical systems begins with the stability 
matrix.  From here, it is possible to calculate 
whether a ray is stable or unstable, what its 
Lyapunov exponent is, and for the unstable ray, 
determine the orientations of the associated 
stable and unstable manifolds that characterize 
the exponential sensitivity to initial conditions.  
All of our results and conclusions are based on 
the behavior of the stability matrices of the 
rays in the model introduced in the previous 
section.  The stability matrix is a strictly local 
analysis in range about some particular reference 
trajectory.  It may be stable at one range, yet 
for a greater range be unstable.  There is no 
restriction that various portions of its full history 
cannot have completely different stability properties.  
In fact, one expects the portions to be almost 
entirely uncorrelated~\cite{prm}.  

Often research done in chaotic 
dynamics uses either time (range) independent or 
periodic Hamiltonians, and the stability matrix is 
investigated about periodic orbits.  The Hamiltonian 
of Eq.~(\ref{hamiltonian}) is aperiodic, and as 
many others have done before, we 
slightly generalize those treatments by considering 
arbitrary, aperiodic rays.

\subsection{Stability Matrix}
\label{stabmat}

The stability matrix for a ray describes the behavior 
of other rays that remain within its infinitesimal 
neighborhood, $\{\delta y,\delta p\}$, for all ranges.  
It is derived by linearizing the dynamics locally; 
see Ref.~\cite{heller} for more details.  At the range 
$r$, one has 

\begin{equation}
\label{stabmatrix}
\left ( 
\begin{array}{c}
\delta p_r \\
\delta y_r 
\end{array}
\right ) 
= M
\left ( 
\begin{array}{c}
\delta p_0 \\
\delta y_0 
\end{array}
\right ) 
\; ,
\end{equation}
with the stability matrix being given by the partial 
derivatives
\begin{equation}
M=
\left (
\begin{array}{cc}
m_{11} & m_{12}\\ m_{21} & m_{22}
\end{array}
\right ) 
=\left (
\begin{array}{cc}
\left. {\partial p_r\over \partial p_0} \right |_{y_0} &
\left. {\partial p_r\over \partial y_0} \right |_{p_0}\\
\left. {\partial y_r\over \partial p_0} \right |_{y_0} &
\left. {\partial y_r\over \partial y_0} \right |_{p_0}
\end{array}
\right ) 
\; .
\end{equation}
The multi-dimensional generalizations are immediate.  The 
$m_{21}$ matrix element is well known for its appearance 
in the prefactor of the standard time (range) Green's function of the 
parabolic equation; it therefore gives directly 
information on wave amplitudes.

The evolution of $M$ is governed by
\begin{equation}
\label{stabev}
\frac{d}{dr} M = KM \; ,
\end{equation}
with the initial condition $M(r=0)$ being the identity matrix, and 
\begin{equation}
K=
\left (
\begin{array}{cc}
-{\partial ^2 H\over \partial y \partial p} & 
-{\partial ^2 H\over \partial y^2} \\
{\partial ^2 H\over \partial p^2} & 
{\partial ^2 H\over \partial y \partial p} 
\end{array}
\right ) 
\Longrightarrow
\left (
\begin{array}{cc}
0 & -{\partial ^2 V\over \partial y^2} \\
1 & 0
\end{array}
\right ) \; .
\end{equation}
The latter form is the simplification relevant for 
Hamiltonians of the so-called mechanical type as in 
Eq.~(\ref{hamiltonian}).  
Since Eq.~(\ref{stabev}) represents
linear, coupled, first-order differential 
equations, the elements of $M$ can be numerically 
calculated as a function of range simultaneously 
with the calculation of its reference ray using identical 
numerical techniques, e.g.~variable step, 
fourth-order Runge-Kutta.

\subsection{Stability and Lyapunov exponents}
\label{lyap}

The stability matrix has several important properties.  
It can be viewed as generating a linear, canonical 
transformation, and therefore its determinant is 
equal to unity.  It is diagonalized by a linear, 
similarity transformation
\begin{equation}
\label{diag}
\Lambda=L M L^{-1}\Longrightarrow \left(\begin{array}{cc}
\lambda & 0 \\ 0 & \lambda^{-1}\end{array}\right) \; ,
\end{equation}
where the last form applies specifically to the case 
of a single degree of freedom.  Here, the second 
eigenvalue must be the inverse of the first in order 
for $\mbox{det}[M]=1$.  The diagonalizing similarity 
transformation leaves the sum  of the diagonal 
elements (trace), $\mbox{Tr}(M)$, invariant.  It is then clear
that $\mbox{Tr}(M)$ is 
real, and three distinct cases may arise.  The 
first is $|\mbox{Tr}(M)|< 2$ which is linked to stable 
motion, and it is then customary to denote 
$\lambda=\exp(i\theta r)$. 
The second case is 
$|\mbox{Tr}(M)|=2$, and it is often called marginally stable 
because it is the boundary case between stable and 
unstable motion.  The third case represents unstable motion, 
and is characterized by $|\mbox{Tr}(M)|> 2$.  In 
Fig.~\ref{random}, 
a typically stable ray is represented by the solid line,
and a typically highly 
unstable ray is represented by the dashed line.  
Their distinctions are not immediately obvious.  

\begin{figure}
\begin{center}
\leavevmode
\epsfxsize = 7.5 cm
\epsfbox{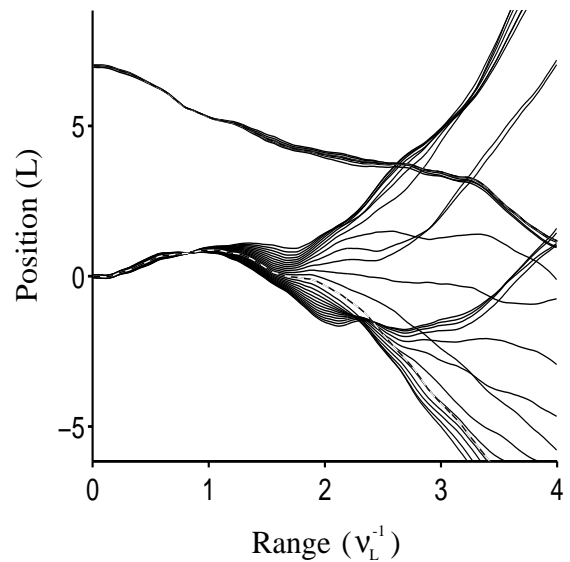}
\end{center}
\caption{Two bundles of rays surrounding the stable
and unstable rays of Fig.~\ref{random} (dashed lines). For the unstable bundle,
12 rays on each side of the reference ray were chosen, each initially
with zero momentum and uniformly sampling the initial position over
a 0.1 L window about the reference ray.  The stable bundle used only
2 rays on each side of the reference ray, but used the same initial
condition domain as for the unstable bundle; this bundle's position
is translated to 7 L.  Note the range scale is in
units of the inverse Lyapunov exponent $\nu_L^{-1}$, defined in
Eq.~(\ref{lyapunov}), and for the potential characterized in Fig.~\ref{vyr},
is approximately 42.94 L.  }
\label{rays}
\end{figure}

The evolution of neighboring rays about a
ray that has $|\mbox{Tr}(M)| < 2$ satisfied from the source to the 
reception range will undergo only rotations in phase space, and
subsets of phase space of finite measure where this 
behavior dominates the dynamics is precisely where 
intermittent-like rays reside.  Fig.~\ref{rays} 
illustrates this characteristic behavior by showing 
a group of stable rays winding about each 
other as they propagate.  The dashed ray in the group 
is the stable ray of Fig.~\ref{random}.  They 
perform their ``random walks'', yet remain winding about 
each other.  For the purposes of this paper, 
we make a slightly generalized definition of intermittent-like 
rays as being all those for which $|\mbox{Tr}(M)|$ remains
sufficiently small over the range of propagation, i.e.~not far from two.  

For unstable motion, it is 
customary to denote $\lambda=\pm \exp(\nu r)$ 
where $\nu$ is positive and real.  
The neighboring rays move 
hyperbolically relative to each other.  
We add a collection of unstable (chaotic) rays onto 
Fig.~\ref{rays} to illustrate the distinction 
between neighboring groups of stable and unstable 
rays.  The dashed ray 
is the highly unstable ray from Fig.~\ref{random}.  
The rays were selected to span the same size 
initial neighborhood as the stable group, 
yet they fan out and become completely independent.  
For the unstable case one can
introduce a definition for the Lyapunov 
exponent as 
\begin{equation}
\label{lyapunov}
\nu_L \equiv \lim_{r\rightarrow\infty} {\ln\left(|\mbox{Tr}(M)|\right)\over r} 
\; .
\end{equation}
Note that there is no ensemble averaging implied in 
the definition of $\nu_L$.  None of the theory presented thus 
far prevents it from taking on a distinct value for 
each ray for each realization of the random potential.  
In Sect.~\ref{fluc}, the
value of $\nu_L$ will be shown to be
independent of the particular ray, and the particular 
realization of the potential as well.  
It thus defines a unique length scale, $\nu_L^{-1}$, which 
is used from Fig.~\ref{rays} onward wherever the 
range variable is involved.  

For unstable motion, $|\lambda|$ tends to be very large 
leaving $\lambda^{-1}$ negligible.  With little 
inaccuracy, $\mbox{Tr}(M)=\lambda+\lambda^{-1} \approx \lambda$ ,
even for finite ranges.  One then deduces a stability 
exponent, $\nu$ from $\mbox{Tr}(M)$ as 
\begin{equation}
\label{nu}
\nu = {\ln|\mbox{Tr}(M)|\over r} \; .
\end{equation}
Thus $\nu$ depends on the particular ray and varies with 
range whereas the Lyapunov exponent has no range 
dependence by definition.  

For any fixed range, an ensemble of $\nu$ can be 
created by considering various initial 
conditions (by exploiting the isotropic and stationary properties
of $V(y;r)$), and different realizations of $V(y;r)$.  
The resulting statistical densities of 
$|\mbox{Tr}(M)|$, $\rho_{|\mbox{Tr}(M)|}(x)$, and similarly 
$\nu$, $\rho_\nu(x)$, are the main objects of concern; the two 
densities are directly tied to each other.  The cumulative 
probability distribution is given as 
\begin{equation}
\label{cpf}
F_\nu(x)=\int^x_{-\infty}\ {\rm d}x^\prime \rho_\nu(x^\prime)
\end{equation}
which provides a useful tool for numerically studying the 
behavior of $\rho_\nu(x)$.  It also has the utility of 
directly giving the proportion of nearly stable rays 
up to some argument set to a maximum instability criterium, $\nu=x$.  

We denote the mean and variance respectively as
\begin{eqnarray}
\label{mean}
\nu_0&=&\langle \nu\rangle = \int\ {\rm d}x\ x\rho_\nu(x) \; ,
\nonumber \\
\sigma^2_\nu&=&\langle\left(\nu-\nu_0\right)^2\rangle=
\int\ {\rm d}x\ \left(x-\nu_0\right)^2\rho_\nu(x) \; ,
\end{eqnarray}
where the brackets $\langle\rangle$ denote ensemble averaging.  
For any real $\gamma$, ensemble averages of powers 
of $|\mbox{Tr}(m)|$ are expressed as
\begin{equation}
\label{trmom}
\langle\left|\mbox{Tr}(M)\right|^\gamma\rangle=
\langle\exp(\gamma\nu r)\rangle=
\int\ {\rm d}x\ \exp(\gamma xr)\rho_\nu(x) \; .
\end{equation}
Note that the case of $\gamma=-1/2$ relates to 
wave amplitude statistics resulting from a semiclassical
reconstruction of the wave field,
and will be discussed in a future work.  

To continue the theoretical development, it is useful 
to introduce a slightly modified stability exponent, $\bar \nu$:
\begin{equation}
\label{lntrm}
\bar \nu \equiv {\ln\langle|\mbox{Tr}(M)|^2\rangle\over 2r} \; .
\end{equation}
Clearly $\bar\nu$ is necessarily greater than $\nu_0$ 
because of the important distinction of 
ensemble averaging before taking the 
natural logarithm as opposed to the inverse 
order and the root mean square fluctuation 
contributions.  
It is shown in the next section that the Lyapunov 
exponent becomes $\nu_L= \lim_{r\rightarrow\infty}\nu_0$, 
and not $\lim_{r\rightarrow\infty}\bar\nu$ which surprisingly 
remains greater than 
$\nu_L$.  Near a parameter regime motivated 
by the ocean, we find numerically
that analytical estimates of $\nu_L$ as being equal to 
$\bar \nu$ are roughly double their actual values.  

\subsection{Stochastic analysis results}
\label{stoch}

An analytic estimate of $\bar \nu$ can be derived 
from previous analytic results based on a 
stochastic analysis which involves a strong 
Markovian assumption~\cite{malakhov,kulkarny,wolfson2}.  
It was verified in Ref.~\cite{wolfson2} that, 
in the context of the present acoustic ray model, 
the stochastic analysis predictions for $\bar \nu$ 
(actually $\nu^\prime$, see text ahead) matched to 
a high degree of precision with numerical tests.  
In fact, no statistically significant deviations 
were observed.  Thus, although the stochastic 
system is not strictly mathematically equivalent 
to the deterministic dynamics, we accept the 
applicability of those specific results at sufficiently long ranges
(defining this range scale is admittedly not as trivial to 
determine for the general ocean acoustics scenario as it 
is for the idealized problem).  
We begin with 
\begin{equation}
\label{tracemsq}
\langle[\mbox{Tr}(M)]^2\rangle=\langle m_{11}^2\rangle 
+\langle m_{22}^2\rangle 
+2\langle m_{11}m_{22}\rangle 
\; .
\end{equation}
By appealing to stochastic integration 
techniques~\cite{papa,dawson,strat}, 
it has been shown that in the 
small-$\epsilon$, large-$r$ limit that ~\cite{malakhov,kulkarny,wolfson2}
\begin{equation}
\label{stocheq}
\langle m_{22}^2\rangle={1\over 3}\exp(2\nu^\prime r)
\; ,
\end{equation}
where 
\begin{eqnarray}
\label{nuprim}
 \nu^\prime & \approx  &
\left( 
{1 \over 2}\int_{0}^{\infty} d\xi 
\left \langle 
\left .
 \frac{\partial^2 V(y;r-\xi)}{\partial y^2}
\right | _{y=y_0 \atop p=p_0}
\right . 
\right . 
\cdot 
 \nonumber \\
 && \hspace{2.5 cm} 
\left . \left . \left . \frac{\partial^2 V(y;r)}{\partial y^2} 
\right | _{y=y_0 \atop p=p_0}
\right \rangle 
\right)^{1/3}   \nonumber \\
&  = & (3\sqrt{\pi})^{1/3}\epsilon^{2/3} 
\end{eqnarray}
(in dimensional units $\nu^\prime=(3\sqrt{\pi})^{1/3} 
\epsilon^{2/3}/L$).  The last result of Eq.~(\ref{nuprim})
is for the specific example of a Gaussian single scale potential
of Eq.~(\ref{gaussian_pot}).  
The first result of Eq.~(\ref{nuprim}) is more general, but requires
numerical confirmation for models with greater 
realism, and will also depend on the ray's initial conditions
for models with a nonuniform background sound speed field.

By symmetry considerations of the stochastic equations, 
$\langle m_{11}^2\rangle = \langle m_{22}^2\rangle$.  
It is also deduced that $\langle m_{11}m_{22}\rangle$ 
can, at most, grow on the same scale.  Defining 
a correlation coefficient, 
\begin{equation}
\label{mucorr}
\mu= {\langle m_{11}m_{22}\rangle \over \langle m_{22}^2\rangle }
\; ,
\end{equation}
where $|\mu|\leq 1$, it follows that
\begin{equation}
\label{trmeval}
\langle[\mbox{Tr}(M)]^2\rangle={2\over 3}(1+\mu)\exp(2\nu^\prime r)
\; .
\end{equation}
Then, by using the definition of Eq.~(\ref{lntrm}), one obtains
\begin{equation}
\bar\nu=\nu^\prime + {1\over 2r} \ln \left[{2\over 3}(1+\mu)\right]
\; .
\end{equation}
Note that the second term disappears if $\mu$ equals $1/2$; 
i.e.~$\langle m_{11}m_{22}\rangle = \langle m_{22}^2\rangle/2$.  
We give its value numerically in the next section.  In that case, 
$\bar\nu=\nu^\prime$ at finite range, and we have an analytic 
estimate for $\bar\nu$ (which has not been derived 
previously to our knowledge).  It is also worth remarking that 
$\nu^\prime$ is not the Lyapunov exponent itself (just as 
$\bar \nu$ is not), but rather only an upper bound.  By 
analogy with the behavior of $\bar\nu$ 
stated at the end of the last subsection, $\nu^\prime$ will turn out 
numerically to be about double the actual $\nu_L$.  

\section{Fluctuations}
\label{fluc}

\begin{figure}
\begin{center}
\leavevmode
\epsfxsize = 7.5 cm
\epsfbox{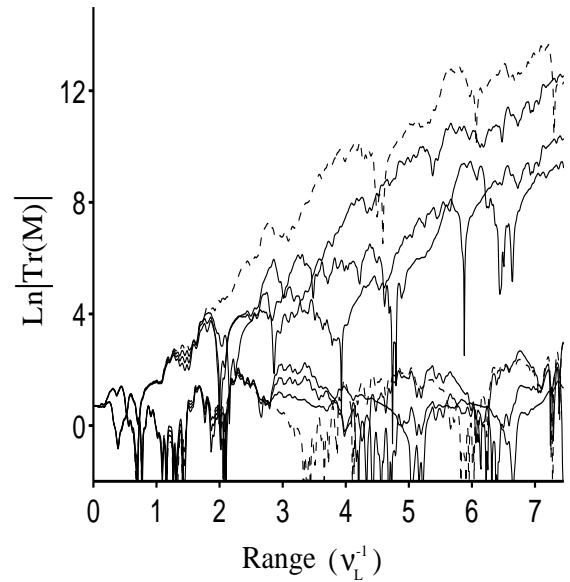}
\end{center}
\caption{The stability, $\ln|Tr(M)|$, for the reference rays of 
Fig.~\ref{random} (dashed), 
and 3 of their neighboring rays initially with zero momentum
and initial position shifted 
$0.008, \; 0.016, \; \mbox{and} \; 0.024 \; \mbox{L}$ 
away from the reference rays. }
\label{nurange}
\end{figure}

The ocean is not infinite in extent, and so the 
distribution of the stability exponents, $\nu$ 
(or $|\mbox{Tr}(M)|$) at a specified range $r$, is more directly 
relevant to the ocean acoustics problem than 
the Lyapunov exponent, $\nu_L$ (or $\exp(\nu_L r)$).  
In order to visualize the magnitude of the 
fluctuations we are discussing, Fig.~\ref{nurange} 
displays the $\ln |\mbox{Tr}(M)|$ for eight of the 
rays from Fig.~\ref{rays} as 
a function of range out to $7.5\nu_L^{-1}$.  
By the right end of the figure, for any fixed 
range one ray might have a $|{\rm Tr}(M)|={\rm e}^{13}$, 
and another one might have $|{\rm Tr}(M)|={\rm e}^{0}$.  
At $7.5\nu_L^{-1}$, there exist fluctuations in the 
stabilities of at least six orders of magnitude 
which is characteristic of broad tailed densities.  

To characterize the fluctuations more 
quantitatively, we consider the cumulative 
densities for $\nu$ and $|\mbox{Tr}(M)|$.  
An initial working hypothesis might be to 
check whether at some long, fixed range, 
a diagonal element of $M$, say $m_{ii}$, is 
distributed as a Gaussian random variable 
across the ensemble of $V(y;r)$ and derive 
the implied cumulative densities from there.  However, 
there ought to be an identifiable mechanism 
for a central limit theorem (CLT) to be operating 
with respect to $m_{ii}$.  From Eq.~(\ref{stabev}), 
one can deduce that $M$ can be decomposed into 
a product of shorter range stability matrices.  
For very long $r$, consider a range $\Delta r$ which is short 
compared to the final range r, yet long compared with $\nu_L^{-1}$.  
Let $N\Delta r=r$ where $N$ is large.  Then it 
follows that the stability matrix is given by 
the left-ordered product
\begin{equation}
\label{product}
M=\prod_{l=1}^N M_l
\; ,
\end{equation}
where $M_l$ is the stability matrix for the range 
$l\Delta r$ to range $(l-1)\Delta r$.  To a great degree of 
accuracy the set of $M_l$ should behave independently with 
the only correlations being amongst the matrix elements 
necessary for maintaining unit determinant.  The stability 
matrix should have the statistical properties of an 
ensemble of products of uncorrelated, random matrices.  

If there exists a limiting form for a distribution 
at long range $r$, one would expect the same 
form (with different parameters, i.e. mean, variance) 
at $r/2$.  In other words, the limiting form would 
have to be invariant under the matrix multiplication 
process.  Denoting $m_{l,ij}$ as the matrix 
elements of $M_l$, for the $N=2$ case, we have 
\begin{equation}
\label{mprod}
m_{11}=m_{2,11}m_{1,11}+m_{2,12}m_{1,21}
\end{equation}
If the $m_{l,ij}$ behave as independent, random Gaussian 
variables, then $m_{11}$ could not be Gaussian 
because of the product form.  The applicability of a 
CLT results from an additive 
process involving random variables.  Instead, we 
anticipate something closer to a lognormal 
density because the log of a product of random 
variables acts like a sum of random variables.
It should be mentioned here that this concept has been
in use in many problems involving statistical 
physics~\cite{prm}.

\begin{figure}
\begin{center}
\leavevmode
\epsfxsize = 7.5 cm
\epsfbox{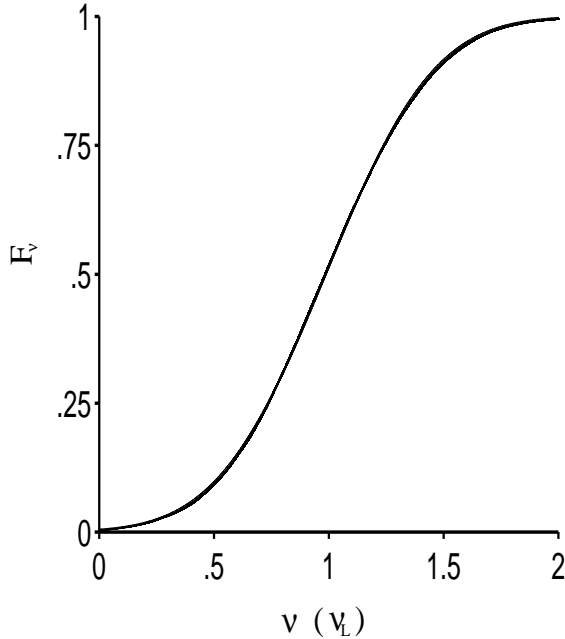}
\end{center}
\caption{Plot of the cumulative density for $\nu$ at the range
of 7.5 $\nu_L^{-1}$. The measured cumulative density is computed from
5 realizations of sound speed fields characterized in Fig.~\ref{vyr} 
($\epsilon = 5\times 10^{-3}$). It incorporates 
20,000 rays per realization whose initial conditions 
uniformly sample 20 L in 
position and have zero initial momentum.  Superposed is the
cumulative density associated with the Gaussian density for $\nu$
[see Eq.~(\ref{nubarnu})] using a value of $\nu_0=0.0232$ (which is our best 
estimate for $\nu_L$ derived from
the same simulations) and $\bar{\nu}=\nu^{\prime}$ [see Eq.~(\ref{nuprim})].
Note the scale for $\nu$ has been normalized by $\nu_L$.}
\label{cumdens}
\end{figure}

\begin{figure}
\begin{center}
\leavevmode
\epsfxsize = 7.5 cm
\epsfbox{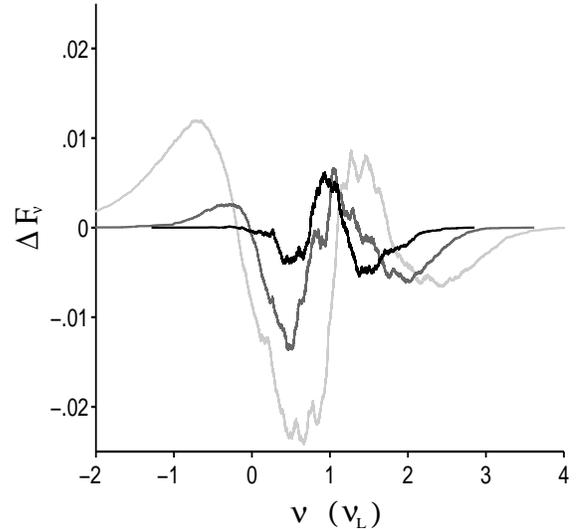}
\end{center}
\caption{Plot of difference between analytical and measured cumulative
densities for $\nu$ 
at the ranges of 1.87 (light gray),3.75 (medium gray), and 7.5 (black) 
$\nu_L^{-1}$.  The measured cumulative densities are taken from 
the same simulations that produce Fig.~\ref{cumdens}.  
The values of the free parameters
$\nu_0$ and $\bar\nu$ were adjusted within their simulated standard deviations
to minimize the maximum difference for each range shown.
Note the scale for $\nu$ has been normalized by $\nu_L$.}
\label{cumdensdiff}
\end{figure}

To test whether $|\mbox{Tr}(M)|$ is lognormally 
distributed, we calculate $100,000$ rays through 
$5$ realizations of $V(y;r)$ to $7.5\nu_L^{-1}$ ($320L$)  (a 
reasonable upper bound for global acoustic 
propagation) for values of 
$\epsilon=2\times 10^{-3} \;,3\times 10^{-3} \;, \mbox{and} \; 
5 \times 10^{-3}$.  
If $|\mbox{Tr}(M)|$ is distributed lognormally, then 
$\nu$ is distributed in a Gaussian manner by definition.  
In Fig.~\ref{cumdens}, we plot the cumulative density 
for $\nu$.  The corresponding analytic Gaussian form 
is superposed.  It is impossible to distinguish the 
numerical results from the Gaussian form from this 
plot.  A similar plot for $|{\rm Tr}(M)|$ carries 
little new information, and is not pictured in this paper, 
though we have verified its excellent consistency 
with a lognormal density as well.  By 
plotting the differences between the numerical and 
analytical curves for three different ranges, we see 
in Fig.~\ref{cumdensdiff} that the consistency with 
a Gaussian density is excellent, and that as range 
increases the consistency of $\rho_\nu(x)$ with a 
Gaussian density improves.  Note the small scale of the 
deviations.  We have verified that they are 
roughly of the order of expected sample size 
errors for the curve at maximum range.  


\begin{figure}
\begin{center}
\leavevmode
\epsfxsize = 7.5 cm
\epsfbox{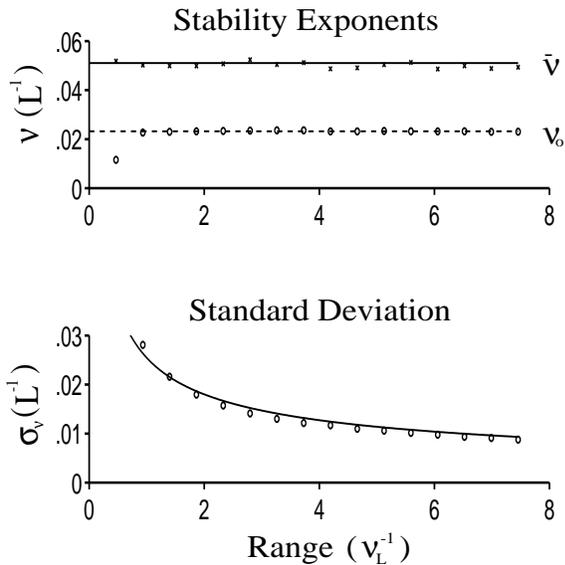}
\end{center}
\caption{Upper panel: The range dependence of $\bar\nu$ derived from
simulations (x's) with the solid line indicating the analytical value
($=0.0510$) determined by Eq.~(\ref{nuprim}).  
Also shown is the range dependence
of $\nu_0$ derived from simulations (o's) with the dashed line indicating
our best estimate of $\langle \nu_0\rangle =0.0232$ 
derived by taking the mean of the 
$\nu_0$ values at ranges beyond 2 $\nu_L^{-1}$.
Lower panel:  The standard deviation of $\nu$ as a function of range
derived from simulations (o's), and the analytical estimate (solid line)
from Eq.~(\ref{sigdep}) using the value $\langle \nu_0 \rangle$ for 
$\nu_0$ (as described in the upper panel) 
and Eq.~(\ref{nuprim}) for $\bar\nu$.}
\label{nunu}
\end{figure}

There are several relationships implied by the 
lognormal density that are straightforward to test.   
First, if we denote the variance of $\nu$ 
as $\sigma_\nu^2$, then a relationship 
between $\bar\nu$ and $\nu_0$ can be derived.  
With 
\begin{eqnarray}
\label{nubarnu}
\rho_\nu(x)&=&{1\over \sqrt{2\pi\sigma^2_\nu}} 
\exp\left[-{(x-\nu_0)^2 \over 2\sigma_\nu^2}\right]
\; , \hspace{2.5 cm}
\nonumber \\
\bar\nu &=& {1\over 2r}\ln \langle e^{2\nu r}\rangle \nonumber \\
&=& {1\over 2r}\ln \left( {1\over \sqrt{2\pi\sigma^2_\nu}}
\int_{-\infty}^\infty {\rm d}x\ \exp (2xr) \cdot \right . \nonumber \\
&& \hspace{2.5 cm} \left .
\exp\left[-{(x-\nu_0)^2 \over 2\sigma_\nu^2}\right]\right)
\nonumber \\
&=& r\sigma^2_\nu + \nu_0 
\; .
\end{eqnarray}
Inverting this last relation for $\sigma^2_\nu$, one obtains
\begin{equation}
\label{sigdep}
\sigma^2_\nu={\bar\nu - \nu_0 \over r} \; .
\end{equation}
Both exponents $\bar\nu,\nu_0$ were defined
(see Eq.~(\ref{nu},\ref{mean},\ref{lntrm}))
to be independent of $r$ to leading order; see the 
upper panel of Fig.~\ref{nunu} where $\bar \nu,\nu_0$ 
are plotted as a function of range.  The stochastic 
approximation for $\nu^\prime$ also given matches precisely 
the value of $\bar\nu$ implying that $\mu=1/2$.  From numerical
simulations, it turns out 
that $\mu$ is 0.466, but this number is poorly determined due to
sample size errors.  There is no 
discernible $r$-dependence in either $\bar\nu$ or 
$\nu_0$ beyond the scale at 
which the stochastic approximation begins to work 
for $\bar\nu$.  They maintain a rather constant ratio of $2.20$ 
amongst themselves.  The lognormal $|\mbox{Tr}(M)|$ density 
thus implies that the standard deviation of 
$\rho_\nu(x)$ approaches zero as $r^{-1/2}$.  Again 
there is excellent consistency; see the lower panel 
of Fig.~\ref{nunu} where $\sigma_\nu$ is plotted 
versus $[(\bar\nu-\nu_0)/r]^{1/2}$.  
Thus, in the limit of $r\rightarrow 
\infty$, $\rho_\nu(x)$ goes to a $\delta$-density;  
all $\nu$ converge to the single value 
$\nu_0$.  This value would also have to be the 
Lyapunov exponent from the definition in 
Eq.~(\ref{lyapunov}), and the Lyapunov exponent would 
be a constant for all trajectories independent of the 
specific realization of the potential or the initial conditions.  
It appears that the approach of $\nu_0$ to $\nu_L$ is so rapid 
as to warrant replacing $\nu_0$ with $\nu_L$ 
in all the formulae of this section.  In fact, 
the lower panel of Fig.~\ref{nunu} actually incorporates our best value 
for $\nu_L$ and not $\nu_0$ as a function of range.  

A consequence of the lognormal density for 
$|\mbox{Tr}(M)|$ is that the density of
$|\mbox{Tr}(M)|^\gamma$ for any real $\gamma$
must also be lognormally distributed.  
This follows from the fact that $\gamma\nu$ would 
be Gaussian distributed with mean $\gamma\nu_0$ 
and variance $\gamma^2\sigma^2_\nu$, and 
$\gamma\nu=\ln|\mbox{Tr}(M)|^\gamma/r$.  Thus $\gamma$ 
enters as a linear scale factor in the parameters 
that define the lognormal density.  It is given 
by
\begin{eqnarray}
\label{rhotrm}
\rho_{|\mbox{Tr}(M)|^\gamma}(x)& = & \sqrt{\frac{1}{2\pi r(\bar\nu-\nu_0)}}
\frac{1}{|\gamma| x}  \cdot \nonumber \\
& & \!\!\!\! \exp{\left [ \frac{- \left 
( \ln{(x)}/\gamma-\nu_0 r \right )^2}{2r(\bar\nu-\nu_0)}
\right ]} 
\; , \; x \geq 0 \; .
\end{eqnarray}
Straightforward integration gives
\begin{equation}
\label{momenttrm}
\langle |\mbox{Tr}(M)|^\gamma \rangle = \exp{
\left(
\left [\gamma \nu_0 + \gamma^2(\bar\nu-\nu_0)/2 \right ] r
\right)} 
\end{equation}
for its ensemble averaged value.  Note that the 
$\gamma=2$ case for which the stochastic theory 
was worked out is the only one independent of $\nu_0$, 
and thus also $\nu_L$ in the large-$r$ limit.  Using 
Eq.~(\ref{momenttrm}), a variety of estimates for the 
Lyapunov exponent can be constructed.  For example, 
for $r$ large enough
\begin{equation}
\label{example1}
\nu_L\approx{2\over r} \ln\langle|{\rm Tr}(M)|\rangle -{1\over 2r} 
\ln\langle|{\rm Tr}(M)|^2\rangle
\end{equation}
as given in~\cite{prm}.  Another example would be
\begin{equation}
\label{example2}
\nu_L\approx{1\over r} \ln\langle|{\rm Tr}(M)|^2\rangle -{1\over 4r} 
\ln\langle|{\rm Tr}(M)|^4\rangle
\end{equation}
etc.

Another interesting, rather curious consequence of the constant 
ratio of $\bar\nu$ to $\nu_0$ is that $\bar \nu$ 
does not approach $\nu_L$ in the 
$r\rightarrow\infty$ limit even though 
$\rho_\nu(x)\rightarrow \delta (x-\nu_0)$.  
Care must be taken to perform the non-commuting 
operations of taking the infinite range limit and ensemble averaging
in the correct order.  Furthermore, the variation of $|\mbox{Tr}(M)|$ 
grows without bound as a function of range $r$, in spite of the 
fact that all the trajectories 
possess equal stability exponents in the limit $r\rightarrow \infty$.
From Eq.~(\ref{momenttrm}), it follows that 
\begin{equation}
\label{blowup}
\sigma^2_{|\mbox{Tr}(M)|}= {\rm e}^{\bar\nu r}\left( 
{\rm e}^{\bar\nu r} - {\rm e}^{\nu_0 r}\right) \sim
{\rm e}^{2\bar\nu r}
\end{equation}
where the last form applies in the large-$r$ limit, 
even though $\sigma^2_\nu$ is approaching zero.  

\begin{figure}
\begin{center}
\leavevmode
\epsfxsize = 7.5 cm
\epsfbox{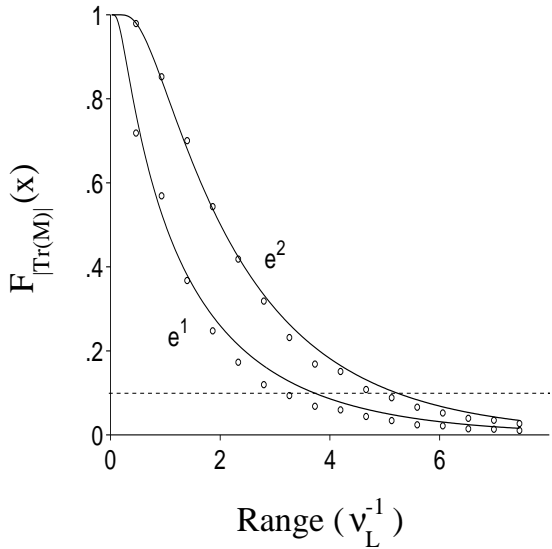}
\end{center}
\caption{The range dependence of the cumulative distribution for $|Tr(M)|$
evaluated at $e^1$ and $e^2$.  The measured values (o's) are derived from
simulations (see caption of Fig.~\ref{cumdens}) 
and the theoretical curves (solid)
are derived from Eq.~(\ref{fracint}) using the values of $\bar\nu$ and
$\nu_0$ indicated in Fig.~\ref{nunu}.  The horizontal dashed line 
at $F_{|Tr(M)|}=0.1$ indicates the range at which $10\%$ of the ray density
remains nearly stable.}
\label{intermit}
\end{figure}

Finally, we point out that a lognormal density has  
long tails and, as already noted, allows for many orders of 
magnitude fluctuations in $|\mbox{Tr}(M)|$.  To return 
to the issue of intermittent-like rays, at any 
range, all rays whose corresponding 
$|\mbox{Tr}(M)|$ are less than some O(1) constant 
can be considered as intermittent-like.  
Values of $e$ or $e^2$ could 
be taken as criteria, for example.  The equivalent 
criteria expressed for the maximum of 
$\nu$ would be $O(r^{-1})$.  In the present model 
the proportion of intermittent-like rays 
approaches zero as $r\rightarrow \infty$, but 
for finite range the proportion of intermittent 
rays up to some maximum value $|\mbox{Tr}(M)|=x$ is 
given by the cumulative density
\begin{eqnarray}
\label{fracint}
F_{|\mbox{Tr}(M)|}(x) &= &F_\nu\left({\ln(x)\over r}\right) \nonumber \\
& = & {1\over 2}\left(1+{\rm erf}\left[ 
{\ln(x)-\nu_0r\over\sqrt{2r(\bar\nu-\nu_0)}}\right] 
\right)
\; ,
\end{eqnarray}
where ${\rm erf}(z)$ is the error function of 
argument $z$.  With the replacement of $\nu_0$ 
by $\nu_L$, this gives a very interesting, nontrivial 
connection between the Lyapunov exponent ($\nu_L$), 
$\bar\nu$, and the proportion of intermittent 
rays as a function of range.  The validity of 
this expression is verified in Fig.~\ref{intermit}.  
The behavior is just as predicted.  The small deviations 
seen may be indicative of some slight non-lognormal 
behavior in the lower tail, or it may just be due 
to our not using best fit values of $\bar\nu$ and $\nu_0$.  
For long ranges, 
the proportion of intermittent-like rays decreases exponentially 
as $a_0 r^{1/2}\exp(-b_0 r)$ where $a_0$ and $b_0$
can be deduced from the asymptotic properties of the error 
function, and this behavior is independent of the 
precise criterium used for the maximum 
desired $|\mbox{Tr}(M)|$.  
As can be seen in Fig.~\ref{intermit}, 
$10\%$ of the initial ray density remains stable or nearly
stable out to ranges of order 
$5 \; \nu_L^{-1}$.  
This $10\%$ of the initial acoustic
energy is then only linearly sensitive to the fluctuating sound speed
field, and since energy remains in coherent bundles 
(see Fig.~\ref{rays}),
they will be expected to have a longer time coherence over repeated
experiments as the environment evolves.  Performing repeated experiments
and applying coherent averaging as a filter to pick up this energy, one can
imagine being able to use this apportionment of the initial acoustic
pulse for acoustic tomography.  

\section{Discussion and Summary}
\label{disc}

Long-range, low-frequency sound propagation in the 
ocean has been previously investigated both as a problem of 
wave propagation through a random medium, and as a 
basis for tomography.  Several outstanding quandaries 
remain that our results only begin to address:  
i) in the early arriving portion of a wave front, 
there seems to be more coherence and stability 
than would be expected from an analysis based on 
stochastic ray techniques common to the subject of WPRM; ii) one 
expects that as one moves from the weak focusing 
to strong focusing regimes (roughly speaking, 
from short range to long range), there should be 
a transition from lognormally distributed wave field 
intensities to Rayleigh distributed ones.  Data 
analyses suggest that the lognormal densities 
extend well beyond the weak focusing regime, and 
the cross-over is not understood theoretically; and 
iii) related to the first item, given the presence 
of more stability than seems consistent with 
theoretical modeling, how valid are the underlying 
assumptions of tomographic inversions performed 
at long ranges?

Complete solutions to these problems are well beyond 
the scope of this paper.  Nevertheless, we believe 
that our results form one cornerstone for their 
eventual resolution.  As the ocean acoustic problem 
mainly involves refraction, and is in a wavelength 
regime that should be extremely well-suited to 
semiclassical analysis, we have focused our 
attention exclusively on a simple ray model 
inspired by the ocean.  Our approach is from a 
dynamical systems perspective as opposed to a 
stochastic ray method.  It has the advantage of being a 
more fundamental starting point in the sense that a 
system's dynamics may determine where a stochastic 
ray method is appropriate, but a stochastic ray 
method just presupposes a certain randomness that may 
or may not actually exist in the system's dynamics.  

Our main concern is the ray stability properties that 
govern wave field amplitudes in semiclassical 
approximations.  A follow-up study is underway 
to address the phases (classical actions and 
geometric indices), correlations amongst ray 
properties, and robustness, i.e.~the generality 
and applicability to the ocean of our results.  
The stability matrix is our key analysis tool 
because it contains all the necessary information 
about how stable or unstable each ray is.  The 
distribution of stabilities reflects on 
statistical properties arising in the study of 
WPRM whereas the existence or lack thereof of 
stable rays impacts tomographic inversion.  We 
also note that studying the stability matrix has 
the utility of providing additional strong checks 
on one's numerical integration techniques.  Its 
determinant must remain unity.  

We have carefully introduced several stability 
exponents depending upon whether ensemble 
averaging is taken before or after the logarithm 
(or at all), and whether the range is finite 
or the infinite range limit is taken.  We have 
related them to the absolute value of the 
trace of the stability matrix 
which we have found to fluctuate to a high degree of 
consistency with a lognormal density; note that this 
also applies to the absolute value of individual matrix elements 
of the stability matrix.  We have 
given a heuristic argument for this distribution, 
and are not aware of any known analytic 
derivations of this result.  

An important consequence follows 
from the appearance of lognormally distributed 
stabilities, or equivalently Gaussian densities in 
the stability exponents 
(the logarithmic variables).  As shown in Sect.~\ref{fluc}
[see Eq.~(\ref{rhotrm})],
any power of the stability 
matrix trace, or individual matrix elements, is also 
distributed lognormally.  Thus, each individual
contribution of an eigenray to the semiclassical 
approximation of the Green's function has a magnitude 
fluctuating as a lognormal density.  Further study 
is underway to determine theoretically the cross-over 
from lognormal wave field intensity distributions 
characteristic of the weak focusing limit to Rayleigh 
densities in the strong focusing limit.  
It is tempting to extrapolate our results to compute
statistically relevant quantities such as the scintillation
index (normalized variance of intensity)
by using Eq.~(\ref{momenttrm}).  Although one
can immediately deduce that the normalized variance of intensity
due to a single ray contribution
grows exponentially with range with a e-folding 
scale of ($\bar\nu -\nu_0$), one cannot infer anything about the
scintillation index in the region where multipathing is important
since the phase of each contributing ray must be incorporated
into the calculation.  Also, our work assumes one
is at or beyond the regime of strong focusing.
This is confirmed in the upper panel of
Fig.~\ref{nunu}, where $\nu_0$ is seen to converge 
at the range $O(\nu_L^{-1})$.
Since the strong Markov assumption is valid for this problem,
this range can be shown to be of the order where strong 
focusing occurs.  Thus it is erroneous to compute the scintillation
index from our work in the weak focusing regime 
($r \ll \nu_L^{-1}$) where only one ray contributes to the 
intensity distribution.  

All rays in the model possess identical Lyapunov 
exponents, and the finite-range, mean stability 
exponent, $\nu_0$, converges rapidly to it (see upper panel of 
Fig.~\ref{nunu}).  This 
follows from the finite range stability exponent 
acting as a Gaussian random variable with a standard 
deviation shrinking with range as $r^{-1/2}$.  This 
is also a consequence of the lognormal density, and 
not the single scale nature of the sound speed 
fluctuations per se.  This behavior may be rather 
general (as general as the lognormal behavior), and 
it would be interesting to verify it in more 
realistic models.  It is quite unlike the 
$\epsilon^{2/3}$-scaling law for the stability 
exponents which should only apply to a model with a 
single correlation scale in range for the sound speed fluctuations.  

The lognormal distribution has very broad tails.  
One typically observes stability matrix traces that 
fluctuate many orders of magnitude at a given range.
Long after the appearance of highly unstable rays 
as a function of range, some stable or nearly 
stable rays will still be present.  Their proportion 
decays essentially exponentially with range where 
the parameters are uniquely fixed by the Lyapunov 
exponent and the related stability exponent $\bar \nu$.  
However, they may be tomographically invertible, and 
relatively speaking, more important than their 
proportion would suggest.  We have pointed out the 
distinctiveness of their behavior relative to 
unstable rays such as the way they twist about each 
other, and hang together as they propagate.  Their 
collective properties appear to be highly correlated.  

The $r^{-1/2}$ behavior of the standard deviation of the 
stability exponent leads to a paradoxical situation in which all the 
rays possess a unique Lyapunov exponent, yet the 
exponentiated quantity, the matrix elements or trace 
of the stability matrix possess a divergent variance as the range 
approaches infinity.  This illustrates dramatically the 
differences arising when non-commuting operations, i.e.~ensemble 
averaging, taking the logarithm, and taking the infinite 
range limit,  are interchanged.  It is the stability 
matrix elements which are relevant to semiclassical 
approximations.  So the individual terms in a summation 
over eigenrays will vary infinitely in their relative 
importance.  \\

{\bf Acknowledgments}

We gratefully acknowledge helpful discussions with M.~G.~Brown, 
F.~D.~Tappert, J.~A.~Colosi, and financial support from the 
Office of Naval Research.

\end{document}